\documentclass[12pt]{emulateapj}

\shorttitle{Star formation to $z>1$ in AEGIS field galaxies}
\shortauthors{Noeske et al.}

\begin{document}



\title{Star formation in AEGIS field galaxies since z=1.1\\ Staged galaxy formation, and 
a model of mass-dependent gas exhaustion}

\author{K.G. Noeske\altaffilmark{1}\email{kai@ucolick.org}}
\author{S.M. Faber\altaffilmark{1}} 
\author{B.J. Weiner\altaffilmark{2}}
\author{D.C. Koo\altaffilmark{1}}
\author{J.R. Primack\altaffilmark{3}} 
\author{A. Dekel\altaffilmark{4}}
\author{C. Papovich\altaffilmark{2}} 
\author{C.J. Conselice\altaffilmark{5}}
\author{E. Le Floc'h\altaffilmark{2}} 
\author{G.H. Rieke\altaffilmark{2}}
\author{A.L. Coil\altaffilmark{2}\altaffilmark{6}}
\author{J.M. Lotz\altaffilmark{7}}
\author{R.S. Somerville\altaffilmark{8}}
\author{K. Bundy\altaffilmark{9}}
%

\altaffiltext{1}{Lick Observatory, University of California, Santa Cruz, CA}
\altaffiltext{2}{Steward Observatory, University of Arizona, Tucson, AZ}
\altaffiltext{3}{Dept.  of Physics, University of California, Santa Cruz, CA}
\altaffiltext{4}{Racah Institute, Hebrew University, Jerusalem, Israel}
\altaffiltext{5}{University of Nottingham, UK}
\altaffiltext{6}{Hubble Fellow}
\altaffiltext{7}{Leo Goldberg Fellow, NOAO, Tucson, AZ}
\altaffiltext{8}{Max-Planck-Institut f\"ur Astronomie, Heidelberg, Germany}
\altaffiltext{9}{California Institute of Technology, Pasadena, CA}


\begin{abstract} 

We analyze star formation (SF) as a function of stellar mass
($M_{\star}$) and redshift $z$ in the All Wavelength Extended Groth
Strip International Survey {\it (AEGIS)}, for star-forming field
galaxies with $M_{\star}\ga10^{10}M_{\odot}$ out to $z=1.1$.
The data indicate that the high specific SF rates (SFR) of many less
massive galaxies do not represent late, irregular or recurrent,
starbursts in evolved galaxies. They rather seem to reflect the onset
(initial burst) of the dominant SF episode of galaxies, after which SF
gradually declines on Gyr timescales to $z=0$ and forms the bulk of a
galaxy's $M_{\star}$.  With decreasing mass, this onset of major SF
shifts to decreasing $z$ for an increasing fraction of galaxies ({\it
staged galaxy formation}). This process may be an important component
of the ``downsizing'' phenomenon.
We find that the predominantly gradual decline of SFR described by Noeske et
al. can be reproduced by exponential SF histories
({\it $\tau$ models}), if less massive galaxies have systematically
longer $e$-folding times $\tau$, and a later onset of SF $(z_f)$. Our
model can provide a first parametrization of SFR as a function of
$M_{\star}$ and $z$, and quantify mass-dependences of $\tau$ and
$z_f$, from direct observations of $M_{\star}$ and SFR up to $z>1$.
The observed evolution of SF in galaxies can plausibly reflect the
dominance of gradual gas exhaustion. The data are also consistent with
the history of cosmological accretion onto Dark Matter halos.
\end{abstract}


\section{Introduction}

In an accompanying Letter (Noeske et al. 2007, hereafter
Paper I), we have studied star formation (SF) rates (SFR) as a
function of stellar mass ($M_{\star}$) and $z$, for field galaxies in
the All Wavelength Extended Groth Strip International Survey (AEGIS)
out to $z=1.1$.  Star-forming galaxies form a defined ``Main
Sequence'' (MS) with a limited range of SFR at a given $M_{\star}$ and
$z$.  This smooth sequence suggests that the same set of few physical
processes governs SF in galaxies, unless quenching of their SF
occurs. The evolution of SF along the MS appears dominated by a
gradual decline of SFR in individual galaxies since $z\sim1$\,, not by
an evolving frequency or amplitude of starbursts.  The dominant
process that governs SF since $z\sim 1$ is hence likely a gradual one,
an obvious candidate being gas exhaustion.

SF histories (SFHs) are known to depend on galaxy mass and
morphological type, both from studies of local galaxies (e.g. Tinsley
1968, Searle et al. 1973, Sandage 1986, Heavens et al. 2004), and from
distant galaxy surveys (see references in \S I of Paper I). The common
picture is that massive galaxies formed the bulk of their stars early
and on shorter timescales, while numerous less massive galaxies evolve
on longer timescales, a phenomenon generally linked to the
``downsizing'' reported by Cowie et al. (1996). 

In this Letter, we show that a simple model of gas exhaustion with
mass-dependent parameters can reproduce and parametrize the observed
SFR as a function of $M_{\star}$ and $z$. Gas exhaustion may thus be
responsible for the gradual decline of SFR that dominates SFHs since
$z\sim 1$ along the MS, i.e. in star-forming field galaxies.
Following previous authors, we consider specific SFR (SSFR),
i.e. SFR/$M_{\star}$, a simple but powerful indicator of galaxy SFHs
(e.g. Kennicutt et al. 2005). We argue that the onset of major SF
occurs systematically later in less massive galaxies.

\section{Data set}   

As in Paper I, we take advantage of the sensitivity and panchromatic
nature of AEGIS; combined SFRs from deep Spitzer MIPS $24\mu$m images and
DEEP2 spectra recover obscured SF in IR-luminous galaxies, and achieve a
large dynamic range in SFR by including galaxies not detected at
$24\mu$m.
For a description of the data, SFR tracers and $M_{\star}$
measurements, see Section 2 of Paper I. We consider all galaxies with
robust SFR tracers a MS galaxy - either $24\mu$m detected, or blue
sequence galaxies with $S/N>2$ emission lines (H$\alpha$, H$\beta$, or
[OII]3727), thereby excluding red LINER/AGN candidates (see Paper I). As
shown in Paper I, this selection likely misses at most $< 10(20)\%$ of
the normally star-forming MS galaxies at $z<(>)0.7$, likely less.

We tested the effects of using different combinations of SFR and
$M_{\star}$ measures, including GALEX UV based SFRs, and $M_{\star}$
values from the color-M/L relation of Bell et al (2003). All
qualitative results of this work are robust against the choice of SFR
tracer or $M_{\star}$ estimate, yet quantitative results will vary
(see \S 4.1).

\section{Parametrization through $\tau$ models}

Interpreting the SSFR vs ($M_{\star}$, $z$) diagrams in terms of
mass-dependent SFH is not straightforward, as $M_{\star}$ grows with
time for SF galaxies. Here we present the use of a simple exponential
model SFH ($\tau$ models, Eq. 1) with mass-dependent parameters to
quantify mass dependences of SF timescales, and to account for
$M_{\star}$ growth.
Previous authors have successfully employed $\tau$ models with
different $e$-folding times $\tau$ to reproduce the spectrophotometric
and chemical evolution of different Hubble types and masses
(e.g. Tinsley 1968, Searle et al. 1973, Koo et al. 1993, Bicker et
al. 2004, Savaglio et al. 2005, Weiner et al. 2006). The apparent
dominance of smoothly declining SFR in individual galaxies (Paper I)
supports the use of $\tau$ models, which are a one-zone approach to
describe SF through continuous gas exhaustion.
We adopt simple closed-box conditions where galaxies have a baryonic
mass $M_b$ that is initially gaseous, later the sum of gas ($M_g$) and
stellar mass $M_{\star}$. For instantaneous recycling, with a
recycled gas fraction $R=0.5$ (Kroupa IMF, Bell et al. 2005), and a SF
efficiency $\epsilon$ such that the SFR $\Psi = \epsilon M_g$, one obtains:
\begin{eqnarray}
\Psi (M_{b},z) = \Psi (z_f) \exp\left(-\frac{T}{\tau}\right),\\
T=t(z)-t(z_f), \tau = \frac{1}{\epsilon(1-R)}
\end{eqnarray}
where $z_f$ is the ``formation redshift'' where SF begins and $t(z)$ is
the cosmic time at redshift $z$
The initial SFR at a given $\tau$ is then $\Psi(z_f)=\epsilon M_b = [\tau(1-R)]^{-1} M_b$. 
We parametrize the mass dependence of $\tau$ as a power law of the
  baryonic mass of the galaxy $M_{b}$:
\begin{equation}
\tau(M_{b}) = c_{\alpha} M_{b}^{\alpha}
\end{equation}
Fig. 1 (left column) shows examples of Eq. 1 in the SSFR-$M_{\star}$
plane, compared to the median SSFR of the MS, for different
$z_f$, $c_{\alpha}$ and $\alpha$.

\begin{figure*}
\plotone{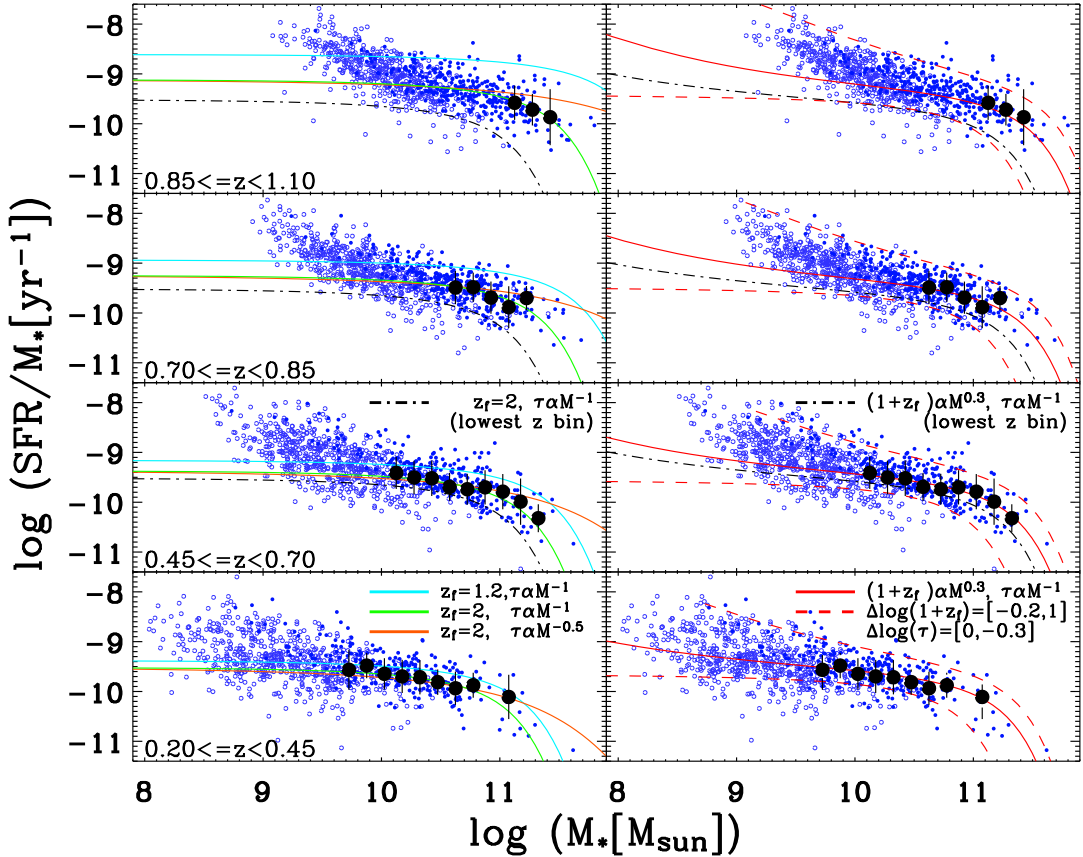}
\caption{Specific SFR $[yr^{-1}]$ vs $M_{\star}$ for $3658$
star-forming (``Main Sequence'', Paper I) AEGIS galaxies.  {\it Full
blue dots}: SFR from Spitzer MIPS 24$\mu$m and DEEP2 emission lines
(Paper I); {\it open blue dots:} blue galaxies without 24$\mu$m
detection, SFR from extinction-corrected emission lines. Galaxies with
no reliable signs of SF, incl. red LINER/AGN candidates (Paper I), are not
shown.  {\it Black dots and error bars:} median and sample standard
deviation of log(SSFR) of the main sequence galaxies, in the
$M_{\star}$ range where the sample is $>95\%$ complete. The {\it black
dot-dashed line} repeats the green (left) and red (right) models in
the lowest $z$ bin. {\bf left:} $\tau$ models with fixed formation
redshift $z_f$ and mass-dependent $\tau$ (colored curves). Massive
galaxies can be reproduced assuming high $z_f(>2)$, less massive
galaxies require $z_f<2$, unphysical for massive galaxies.
{\bf right:} {\it Staged} $\tau$ models (red), where both $\tau$ and
$z_f$ are mass-dependent. Red dashed lines show the effect of varying
$z_f$ and $\tau$ at a given $M_{bar}$.  The delayed onset of SF (lower
$z_f$) in a fraction of less massive galaxies accounts for the
increase of specific SFR at low $M_{\star}$ without requiring a large
fraction of galaxies to simultaneously undergo starbursts.}
\end{figure*}

\subsection{Staged $\tau$ models}

Fig. 1 (left column) shows that models with mass-dependent $\tau$ can
crudely reproduce the median MS of SF galaxies and its redshift
evolution for galaxies with $M_{\star}\ga 10^{10}M_{\odot}$ out to
$z\sim1$, if formation redshifts $z_f\sim 2$ are adopted for all
galaxies.
However, the models fall short of reproducing the high SSFR of less
massive galaxies. The model SSFR remain systematically too low, unless we
adopt a very low $z_f<1$, unphysical for massive galaxies. 
The reason is the monotonic decline of the SFR of $\tau$ models. Their
present-to-past average SFR (Kennicutt et al. 2005),
\begin{equation}
b(t) = \frac{\Psi(t)}{<\Psi >_{T}}=\frac{\Psi}{M_{\star}}\frac{T}{1-R},
\end{equation} 
is always $<1$. The limit for $\tau=\infty$ is $b=1$, which
corresponds to the constant SFR that would have formed a galaxy's
$M_{\star}$ since $z_f$. Empirically, the behavior of the MS suggests
declining SFHs ($b<1$), which causes a conflict between the high SSFR for low
mass galaxies and the assumption that all galaxies started forming
stars at high $z_f (\ga 3)$. Adopting such high $z_f$, an early start
of SF, implies low past-average SFR: $b=1$ then corresponds to low
SSFR, reflected in the low upper limits to the SSFR of $\tau$
models. For these $z_f$, the high SSFR of many less massive galaxies
would imply $b>1$, a current SFR above the past-average level, i.e. an
episode of enhanced SF (Fig. 1, left; Fig. 2).

The increase of the highest SSFR towards less massive galaxies can be
reproduced by allowing the onset of SF to be delayed to lower $z_f$ in
less massive galaxies (see the models for different $z_f$ in Fig. 1,
left). We parametrize $z_f$ as a function of mass, similarly to
$\tau$, an approach we refer to as {\em staged} $\tau$ models.
\begin{equation}
1+z_f(M_{b}) = c_{\beta} M_{b}^{\beta}
\end{equation}
This model interprets high SSFR as the early epoch of smooth SFHs with
lower $z_f$, rather than late episodes of enhanced SF. It is physically
motivated, not an attempt to force an oversimplified model to fit
complex SFHs (see \S 4.2).
Staged $\tau$ models (right column in Fig. 1) provide a better
description of the median of the MS than $\tau$ models with fixed
$z_f$ in the $M_{\star}$ range where the sample is complete, and
appear to describe the data also towards lower $M_{\star}$. Staged
models that consider a moderate range of $z_f$ and $\tau$ at a given
mass (dashed red curves) also reproduce the upper envelope of the MS,
which is complete at all observed $M_{\star}$ and $z$, and the lower
envelope and apparent broadening of the MS towards lower masses.
Models with a range of $z_f$ but no trend with mass would merely
introduce a scatter and an offset in the asymptotic SSFR at low
masses, but would not change the slope of SSFR($M_{\star}$) to first
order (see the models for different $z_f$ in Fig. 1, left column).

The staged $\tau$ model in Fig. 1 (right) is given by
\begin{equation}
\tau=10^{20.7}\left(\frac{M_b}{M_{\odot}}\right)^{-1}yr, 
(1+z_f) = 10^{-2.7}\left(\frac{M_b}{M_{\odot}}\right)^{0.3}.
\label{best}
\end{equation} 
We calculated $\chi ^2$ to scan the model parameter space, but
 Eq. \ref{best} is hand-adjusted to reproduce the median and upper and
 lower limits of the data. Results of a simple $\chi ^2$ minimization
 to the median would be misleading: best values for all four
 parameters in Eqs. (3) and (5) depend considerably on systematics of,
 e.g., SFR and $M_{\star}$ estimates, and the IMF; also, the $\tau(M_b)$
 dependence is mainly constrained by massive galaxies (Fig. 1) where
 our number statistics are poor, and the $z_f(M_b)$ relation at low
 masses, where data are incomplete. An evaluation of the relevant
 uncertainties must incorporate scatter in $\tau(M_b)$ and
 $z_f(M_b)$, or a scatter about smooth SF histories, and is postponed
 to a forthcoming paper.

\section{Discussion}

\subsection{$\tau$ models, gradual decline of SF}

By direct measurement of SFR and $M_{\star}$ over a large range in
mass and $z$, we confirm that the commonly adopted exponential model
SFHs can reproduce the average SFH of MS galaxies. This model can
quantify the mass dependence of the associated SF timescales $\tau$
and of the $z_f$. The mass dependences of $\tau$ and $z_f$, and
$M_{\star}$ growth through SF, conspire to reproduce the decline of
SFR that is similar over a wide $M_{\star}$ range (Paper I; Zheng et
al. 2007).

Notably, $\tau$ models are a simple approximation of SF that declines
due to gradual gas exhaustion. Their ability to reproduce the
evolution of the MS of SF galaxies, along with the limited range of SFR
on the MS, implies that gradual gas exhaustion with mass-dependent
timescales is a plausible driver of the dominant evolution of SF in
galaxies $\ga 10^{10}M_{\odot}$ since $z\sim 1$.
%

We chose a closed box model which is sufficient to reproduce the
coevolution of SF and $M_{\star}$. 
Linking the model $M_{\star}$ to, e.g. Dark Matter halo masses should
involve the observed relation between both values at a given $z$, to
account for gas accretion and removal in galaxies, which are both not
well understood. The $\tau$ models' similarity to
the data does {\em not} imply a closed box scenario where gas is
merely turned into stars. Additional processes that gradually deplete
cold gas --- heating or loss --- and scale roughly with SFR would also
produce SFHs that resemble exponentials.  These processes include
feedback from SF, and conceivably AGN, given the likely co-evolution
of stellar bulges and Black Holes (e.g. Granato et al. 2004). Short
$\tau$ obtained for massive galaxies may largely reflect such gas loss
processes rather than very efficient SF.

We have considered the depletion of an existing gas reservoir, but the
decline of SFR at a given mass is also compatible with the
cosmological decline in accretion onto dark halos.  This can be
approximated for halo masses near $10^{12}M_{\odot}$ by
$\dot{M_h}/M_h\approx 0.04\,{\rm Gyr}^{-1}\,(1+z)^{2.25}$ (Birnboim et
al.  2007), giving a factor of $\sim 5$ between $z\sim 1$ and $z\sim
0$, similar to the observed decline.
The mean virial accretion in this mass range is predicted to vary as
$\dot{M_h}\propto M_h^{1.15}$. If we adopt $M_{\star}/M_h \propto V^2
\propto M_h^{2/3}$ (Dekel \& Woo 2003 for SN feedback), and, naively,
$\dot{M_*} \propto \dot{M_h}$, we obtain $\dot{M_{\star}} \propto
M_{\star}^{0.69}$, compatible with the observed mass dependence (Paper
I).

\subsection{Staged galaxy formation}

\begin{figure}
\plotone{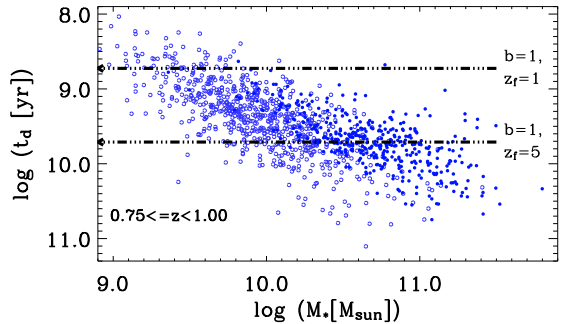}
\caption{Stellar mass doubling times $t_d$ (see \S 4.2) as a function
of $M_{\star}$. Symbols are as in Figure 1. Horizontal lines give the
time between $z_f$ and the center of the $z$ bin shown, for $z_f$ of 1
and 5. Galaxies above these lines would have $t_d$ shorter than their
adopted age, i.e. SFR above their past-average level ($b>1$). Galaxies
on the lines form stars at their past-average rate ($b=1$).}
\end{figure}

Fig. 1 shows the previously described increase of the highest SSFR
towards lower $M_{\star}$. High SSFR have been interpreted
as episodes of SFR above the past average level ($b>1$), based on the
assumption that all galaxies had a relatively high $z_f$ (see \S\,3)
(e.g., Bell et al. 2005, Juneau et al. 2005, Feulner et al. 2005).

Our data indicate that high SSFR often do not represent such irregular
or periodic episodes late in the SFH of a galaxy.  In Fig. 2, we show
SSFR vs $M_{\star}$ as in Fig. 1. We express SSFR in terms of the
{\it doubling times},
$t_d=M_{\star}/[(1-R)\Psi] = [(1-R)\,SSFR]^{-1}$, within which the
current SFR would produce the observed $M_{\star}$. 
Small $t_d$ corresponds t high SSFR; for a declining SFH, a galaxy
can be at most as old as $t_d$.  Consider the galaxies at $0.75<z<1$
and $10^{10} M_{\odot}$, where the data are $>80\%$ complete.
Incompleteness affects mostly red galaxies, hence the plot shows
$>80\%$ of the SF galaxies.  Their current SFR would generate their
observed $M_{\star}$ within $t_d$ of $6\times 10^8$ to $6\times10^9$,
yr, mostly $< 3\times 10^9$\,yr.  These $t_d$ are smaller than half of
the age of the Universe at $0.75<z<1$. If we assume a high $z_f(\sim
5)$, $85\%$ of the galaxies have a SFR above the past average $(b>1)$,
and $57\%$ a starburst ($b>2$, Kennicutt et al 2005)
\footnote{Even if all our diagnostics overestimated SFR by a factor of
2, the fraction of galaxies with $b>1(2)$ would still be 57(21)\%.}.
At face value, this is contradictory; it should not be possible
for a majority of galaxies to simultaneously undergo a starburst, which 
by definition should occupy a short part of a duty cycle.

However, these galaxies form the MS, and their SFR show a gradual
decline on Gyr timescales since $z\sim 1$ (Paper I), likely even since
$z=1.4$, {\em not} an enhancement of SFR in the $z$ range shown in
Fig. 2. These galaxies therefore do not seem to simultaneously undergo
a brief ($\la 1$\,Gyr) episode of elevated SFR on top of lower-level
SFHs; instead, their high SSFR represent the early, strongly
star-forming phase of a SFH that smoothly declines on Gyr timescales
to $z=0$. Their high SSFR (short $t_d$) imply (i) that this gradually
declining epoch must form the bulk of their $M_{\star}$, and (ii) that
the galaxies must be observed $\la 1 t_d$ after the onset of this
epoch, suggesting $z_f\la 2$ for $>60\%$ of these galaxies; otherwise
the produced $M_{\star}$ would be higher than observed.

These lines of evidence suggest that the observed high SSFR of many
galaxies are not due to a periodic or irregular burst, late in their
SFH. Instead, many such galaxies seem to be observed shortly after
their ``initial burst'' phase, the early stage in their predominantly
smooth SFH which forms most of their $M_{\star}$.  Moreover, the
average SSFR increases smoothly to lower masses, at all $z$. This
points to a smooth dependence of the average $z_f$ on galaxy
mass. Based on this evidence, we propose a scenario of {\it ``staged
galaxy formation''}, where the {\em average} onset of the major SF
($z_f$) decreases smoothly with galaxy mass. 
This scenario achieves high SSFR without requiring that a large
fraction of galaxies at any epoch are elevated in SFR ($b>1$) or
starbursting ($b>2$).  Allowing lower $z_f$ for a fraction of less
massive galaxies is the only possibility to avoid this contradiction
between burst fraction and duty cycle.
The staged $\tau $ models we use to approximate these SFHs parametrize
both the decline timescales ($\tau$), and the onset ($z_f$), of the
main SF episode as a function of mass.

The range of $t_d$ in Fig. 2 shows that the staged scenario only requires
a fraction of less massive galaxies to form later: the range of
$z_f$ must reach to lower $z$ for less massive galaxies.  In addition,
the model does not exclude some low level SF prior to the onset of the
major SF episode, effectively the epoch of assembly.  This allows it
to be consistent with the presence of old ($\sim 10$\,\,Gyr) stars in
many low mass local galaxies (e.g. Grebel 2004).

A relation between the galaxy mass and the onset time of the dominant
SF episode is observationally and theoretically supported: see Heavens
et al. (2004); Iwata et al. (2007); Thomas et al. (2005) for
early-type galaxies; Feulner et al. (2005), Reddy et al. (2006) and
Erb et al. (2006) report a systematic decrease of stellar age with
$M_{\star}$ up to $z=3-5$. CDM structure formation provides a
framework for a systematic relation between the dominant SF epoch and
present-day galaxy mass (Neistein et al. 2006).
Finally, insofar as "downsizing" means that a characteristic epoch of
high SSFR occurs early in high mass galaxies, while at low $z$ only low
mass galaxies exhibit high SSFR (Cowie et al 1996), a delayed onset of
major SF in less massive galaxies is a natural part of this process.

\acknowledgements See the survey summary paper (Davis et al. 2006,
this volume) for full acknowledgments.  This work is based on
observations with the W.M. Keck Telescope, the Hubble Space Telescope,
the Galaxy Evolution Explorer, the Canada France Hawaii Telescope, and
the Palomar Observatory, and was supported by NASA and NSF grants.
This work is based in part on observations made with the {\it Spitzer
Space Telescope\/}, which is operated by the Jet Propulsion
Laboratory, California Institute of Technology under a contract with
NASA.  Support for this work was provided by NASA through contract
numbers 1256790, 960785, and 1255094 issued by JPL/Caltech. KGN
acknowledges support from the Aspen Center for Physics. We wish to
thank the referee for helpful comments.

\end{document}